\documentclass[fleqn,12pt,twoside]{article}
\usepackage{espcrc1}
\usepackage{graphicx}

\newcommand{\be}{\begin{equation}}
\newcommand{\ee}{\end{equation}}
\newcommand{\bea}{\begin{eqnarray}}
\newcommand{\eea}{\end{eqnarray}}

\newcommand{\g}{\gamma}
\newcommand{\f}{\frac}

\newcommand{\bra}{\langle}
\newcommand{\ket}{\rangle}

\newcommand\lr[1]{{\left({#1}\right)}}

\title{Gluon production and diffraction in the dipole picture}

\author{C. Marquet\address{Service de Physique Th\'{e}orique, 
        CEA/Saclay, \\ 91191 Gif-sur-Yvette cedex, France}}
       
\begin{document}

\maketitle

\begin{abstract}
Using the eikonal approximation, we show that inclusive gluon production is 
related to the scattering of a gluon-gluon dipole while diffractive gluon 
production in DIS is related to a two-$q\bar q$ dipole scattering amplitude. 
Hence diffractive photon dissociation cross-sections are observables that  
provide potential information on dipole correlations, which represent an open 
issue in high-energy QCD.

\end{abstract}

\section{INTRODUCTION}

Over the past ten years, the perturbative $q\bar q$ dipole picture has been 
developed~\cite{dipole} with the aim of understanding high-energy scattering in 
QCD. This formalism is well-suited to study scattering near the unitarity limit 
because density effects and non-linearities that lead to saturation and 
unitarization of the scattering amplitudes can be taken into account. The dipole 
picture has had great succes for the phenomenology of hard processes initiated 
by virtual photons, in which case the link between the experimental probes and 
the scattering of $q\bar q$ dipoles is established. This is not the case for 
hadron-hadron collisions in which the probes are the final-state particules that 
one measures, {\it e.g.} a final-state gluon that we detect as a jet. 
Establishing a link between the scattering of dipoles and observables 
like jet cross-sections is of great theoretical and phenomenological 
interest for the Tevatron and in the prospect of the LHC.

In the purpose of doing so, we derive the cross-sections for inclusive and 
diffractive gluon production in the high-energy scattering of a $q\bar q$ dipole 
off an arbitrary target and show that they are related to the scattering of 
dipole configurations. In the inclusive case, the relevant object is a 
gluon-gluon dipole ($gg$ dipole) and one can show this is the case independently 
of the incident projectile. In the diffractive case we find that two-$q\bar q$ 
dipole amplitudes are involved; this provides a link between observables like 
diffractive photon dissociation cross-sections measurable at HERA and 
correlations between dipoles. 

\section{HIGH-ENERGY EIKONAL SCATTERING}

Let us start with the eikonal approximation for quarks and gluons scattering at 
high energies. When a system of partons propagating at nearly the speed of light 
passes through a target and interacts with its gauge fields, the dominant 
couplings are eikonal: the partons have frozen transverse coordinates and the 
gluon fields of the target do not vary during the interaction. This is justified 
since the time of propagation through the target is much shorter than the 
natural time scale on which the target fields vary. The effect of the 
interaction with the target is that the partonic components of the incident 
wavefunction pick up eikonal phases: if $|(\alpha,x)\ket$ (resp. $|(a,x)\ket$) 
is the wavefunction of an incoming quark of color $\alpha\!\in\![1,N_c]$ (resp. 
gluon of color $a\!\in\![1,N^2_c\!-\!1]$) and transverse position $x$ (the 
irrelevant degrees of freedom like spins or polarizations are not explicitly 
mentioned), then the action of the ${\cal S}-$matrix is (see for 
example~\cite{eikap}):
\be
{\cal S}|(\alpha,x)\ket\!\otimes\!|t\ket=
\sum_{\alpha'}\left[W_F(x)\right]_{\alpha\alpha'}
|(\alpha',x)\ket\!\otimes\!|t\ket\ ,\hspace{0.3cm}{\cal 
S}|(a,x)\ket\!\otimes\!|t\ket=\sum_{b}W_A^{ab}(x)
|(b,x)\ket\!\otimes\!|t\ket\ ,\label{smat}\ee
where $|t\ket$ denotes the initial state of the target. The phase shifts due to 
the interaction are the color matrices $W_F$ and $W_A,$ the eikonal 
Wilson lines in the fundamental and adjoint representations respectively, 
corresponding to propagating quarks and gluons. They are given by
\begin{equation}
W_{F,A}(x)={\cal P}\exp\{ig_s\int dz_+T_{F,A}^a{\cal A}_-^a(x,z_+)\}
\label{wils}\end{equation}
with ${\cal A}_-$ the gauge field of the target and $T_{F,A}^a$ the 
generators of $SU(N_c)$ in the fundamental ($F$) or adjoint ($A$)
representations. We use the light-cone gauge $\cal{A}_+\!=$0 and
${\cal P}$ denotes an ordering in the light-cone variable $z_+$ 
along which the incoming partons are propagating.

For an incoming state $|\Psi_{in}\ket,$ the outgoing state 
$|\Psi_{out}\ket\!=\!{\cal S}|\Psi_{in}\ket\!\otimes\!|t\ket$ emerging from the 
eikonal interaction is obtained by the action of the 
${\cal S}-$matrix on the partonic components of $|\Psi_{in}\ket$ as indicated by 
formula (\ref{smat}). The outgoing wavefunction $|\Psi_{out}\ket$ is therefore a 
function of the Wilson lines (\ref{wils}). When calculating physical observables 
from $|\Psi_{out}\ket$, one obtains objects that are target averages of traces 
of Wilson lines (the traces come from the color summations that one has to carry 
out). As an example, the simplest of these objects is
\be 
S_{q\bar q}(x,x')=
\f1{N_c}\left\bra\mbox{Tr}\lr{W_F^\dagger(x')W_F(x)}\right\ket_t\  
\label{qqdip}\ee
where we have denoted the target averages $\bra t|\ .\ |t\ket\!=\!\bra\ .\ 
\ket_t.$ This is the $q\bar q$ dipole elastic ${\cal S}-$matrix ($x,$ $x':$ 
positions of the quark and antiquark) which enters for example in the DIS total 
cross-section; more generally, observables are functions of (\ref{qqdip}) or 
more complicated amplitudes. To compute these amplitudes, one has to evaluate 
the averages $\bra\ .\ \ket_t$ which amounts to calculating averages of Wilson 
lines in the target wavefunction. A lot of studies are devoted to this problem, 
here we only establish the link between the observables and the dipole 
amplitudes as we shall do now with gluon-production cross-sections.

\section{INCLUSIVE AND DIFFRACTIVE GLUON PRODUCTION}

\begin{figure}[htb]
\begin{minipage}[t]{85mm}
\includegraphics[width=8cm]{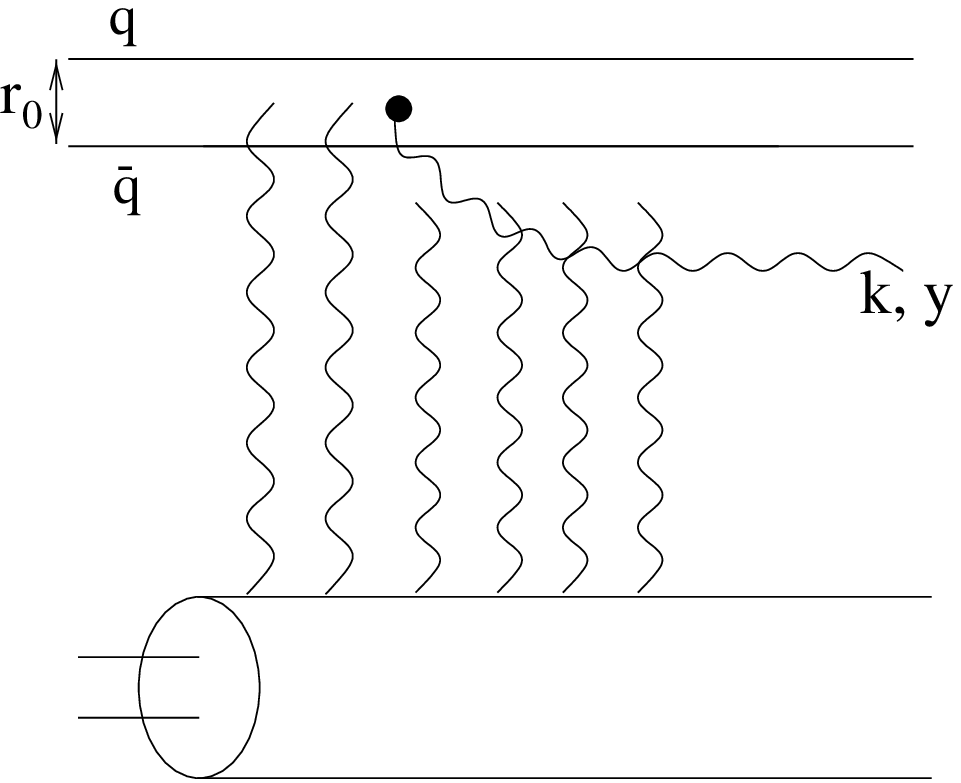}
\caption{Gluon production off a $q\bar q$ dipole of size $r_0.$ $k$ and $y:$ 
transverse momentum and rapidity of the measured gluon.}
\label{gluprod}
\end{minipage}
\hspace{\fill}
\begin{minipage}[t]{70mm}
\includegraphics[width=6.5cm]{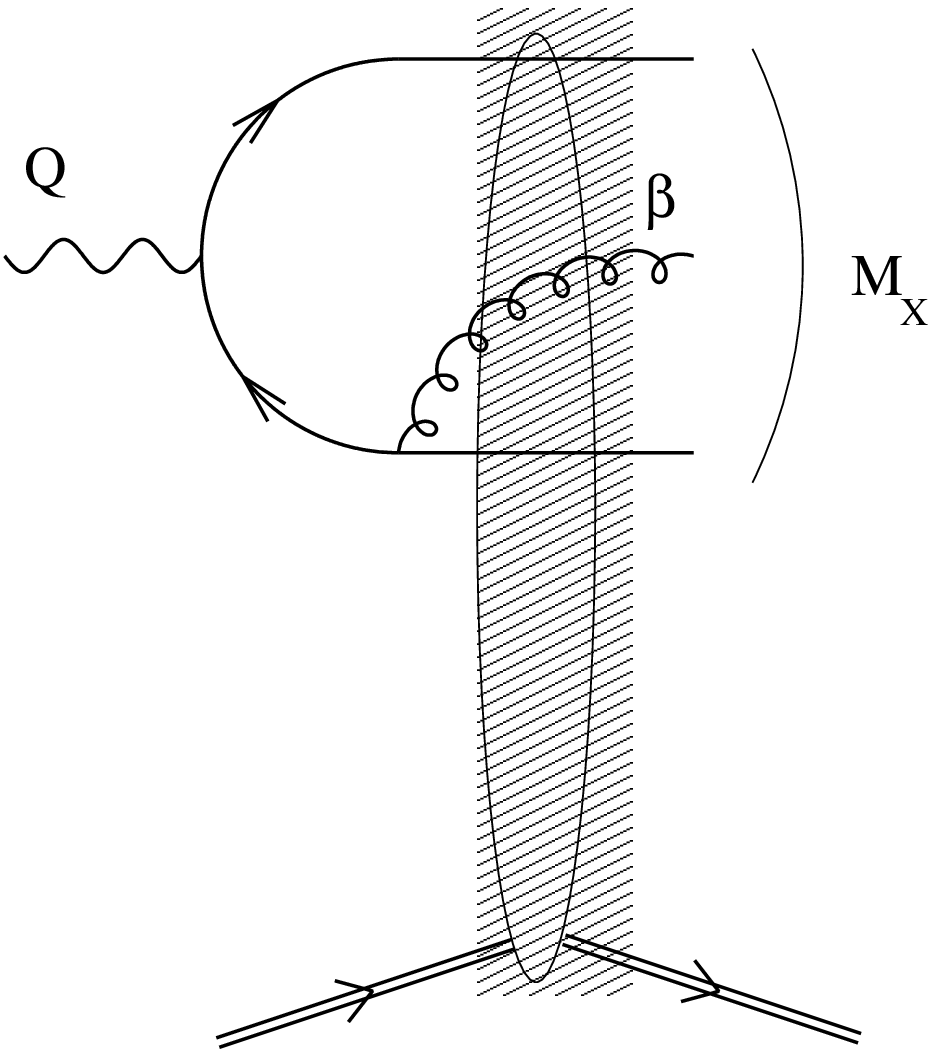}
\caption{Diffractive dissociation of a photon of virtuality $Q.$ $M_X:$ mass of 
the final state. $y\!=\!\log{1\!/\!\beta}.$}
\label{diffdis}
\end{minipage}
\end{figure}

Let us consider that the incoming state is an incident $q\bar q$ dipole of 
transverse size $r_0$ ($|r_0|\!\ll\! 1/\Lambda_{QCD}$ in order to justify the 
use of perturbation theory), see Figure~\ref{gluprod}. The inclusive 
cross-section for the production of a gluon of transverse momentum $k$ and 
rapidity $y$ in the scattering of this $q\bar q$ dipole off an arbitrary target 
then reads~\cite{marq}
\be 
\f{d\sigma_{incl}}{d^2kdy}(r_0)=\f{4\alpha_sC_F}{\pi k^2}
\int_0^{|r_0|} dz\ J_0(kz)\log{\f{|r_0|}z}\ \f{\partial}{\partial z}z
\f{\partial}{\partial z}\int d^2b\lr{1-S_{gg}\lr{b\!+\!\f{z}2,b\!-\!\f{z}2}}\ 
,\label{simp}\ee
where $S_{gg}$ is ${\cal S}-$matrix for the scattering of a 
$gg$ dipole on the target and is given by
\be
S_{gg}(x,x')=\f1{N_c^2\!-\!1}\left\bra\mbox{Tr}
\lr{W_A^\dagger(x')W_A(x)}\right\ket_t
\ .\label{csection}\ee
The azimutal angle of the incoming $q\bar q$ dipole has been integrated out so 
the cross-section depends only on $|r_0|.$
The diffractive cross-section for gluon production reads~\cite{marq} (we define 
the diffractive process as one in which a color singlet is exchanged and the 
target does not break up):
\be
\f{d\sigma_{diff}}{d^2kdy}(r_0)=\f{\alpha_sN_c^2}{4\pi^2C_F}\int d^2b\
{\bf A}\lr{k,b\!+\!\f{r_0}2,b\!-\!\f{r_0}2}.
{\bf A}^*\lr{k,b\!+\!\f{r_0}2,b\!-\!\f{r_0}2}\label{diff}
\ee
where the two dimensional vector ${\bf A}$ is given by
\be
{\bf A}(k,x,x')=\int\f{d^2z}{2\pi}\ e^{-ik.z}
\left[\f{z\!-\!x}{|z\!-\!x|^2}-\f{z\!-\!x'}{|z\!-\!x'|^2}\right]
\lr{S^{(2)}_{q\bar q}(x,z;z,x')-S_{q\bar q}(x,x')}.\ee
\be
S^{(2)}_{q\bar q}(x,z;z',x')=
\f1{N_c^2}\left\bra\mbox{Tr}\lr{W_F^\dagger(x')W_F(z')}
\mbox{Tr}\lr{W_F^\dagger(z)W_F(x)}\right\ket_t\
\ee
is the ${\cal S}-$matrix for the scattering of two $q\bar q$ dipoles on the 
target. To write down these cross-sections, it is assumed that the measured 
gluon is soft, that is we are working in the leading logarithmic approximation 
in $1\!/\!\beta\!\equiv\!e^y.$ The ${\cal S}-$matrices involved in those 
formulae contain the 
scatterings with all numbers of gluon exchanges with the target and, via the 
quantum evolution of the target, they also contain the emissions of gluons 
softer than the measured one $(k,y).$ Then if $Y$ is the total rapidity, the 
${\cal S}-$matrices depend on $Y\!-\!y.$

The result for the inclusive cross-section (\ref{simp}) can be extended to an 
arbitrary projectile~\cite{marq} if its natural scale $1\!/r_0$ is much smaller 
than $k,$ the result being valid at double leading logarithmic accuracy. This is 
of great importance in the prospect of jet cross-sections at the LHC. For the 
diffractive cross-section (\ref{diff}), the same extension does not appear to be 
possible, this could be related to the breakdown of collinear factorization for 
diffractive cross-sections.

\section{DIFFRACTIVE PHOTON DISSOCIATION}

The most straightforward application to our result for the diffractive 
cross-section is the diffractive photon dissociation at large mass in DIS: a 
photon of virtuality $Q^2$ scatters on a target proton which stays intact, see 
Figure~\ref{diffdis}. If the final-state diffractive mass is large enough, the 
cross-section is dominated by the $q\bar qg$ component and is given by our 
result convoluted with the photon-to-$q\bar q$ dipole wavefunction $\phi^{\g}$:
\be
\f{d\sigma_{diff}^\g}{d^2k\ dM_X}=\f{2}{M_X}
\int d^2r_0\
\phi^{\g}(|r_0|,Q^2)\ \f{d\sigma_{diff}}{d^2kdy}(r_0)\ .\label{difphodis}\ee
Phenomenology has been done~\cite{munsho} for the cross-section integrated over 
$k^2$ neglecting the correlations (i.e. writting $S^{(2)}_{q\bar 
q}\!=\!S^2_{q\bar q}$). Taking them 
into account might be a difficult task but seems necessary in order to gain 
insight on dipole correlations from the observable (\ref{difphodis}) and the 
related integrated cross-sections.

As an example, the recent treatment of correlations in~\cite{dipcor} allows a 
first application. Neglecting non-dipole amplitudes in the complicated system of 
equations verified by the ${\cal S}-$matrices in the high-energy limit, the 
authors found a reduction of the problem: $1\!-\!S^{(2)}_{q\bar 
q}(x,z;z',x')\!=\!c\lr{N(x,z)\!+\!N(z',x')\!-\!N(x,z)N(z',x')}$ and 
$1\!-\!S_{q\bar q}(x,x')\!=\!cN(x,x')$ where $0\!<\!c\!<\!1$ is an arbitrary 
number mesuring the strength of the correlations and where $N$ is solution of 
a closed equation called the BK equation~\cite{bk}. One has then for example
\be
\sigma_{diff}^\g=c^2\left.\sigma_{diff}^\g\right|_{BK}\ee
where $c$ is put to 1 in $\sigma_{diff}^\g|_{BK}.$ This quantity is fully 
computable from a solution of the BK equation. The value of $c$ could then be 
extracted from experiments.

Precise measurements of $\sigma_{diff}^\g$ should be made to try and take 
advantage of this link between large-mass diffraction and dipole correlations 
while deeper theoretical studies are being carried out.

\end{document}